\documentclass[runningheads]{llncs}
\usepackage[T1]{fontenc}
\usepackage{graphicx,verbatim}
\usepackage{amsmath,multicol,multirow,amssymb,hyperref,booktabs,makecell}
\usepackage{color}

\begin{document}
\title{GyralNet Subnetwork Partitioning via Differentiable Spectral Modularity Optimization}
\titlerunning{ Spectral Modularity Optimization for GyralNet Subnetwork Partitioning}                                         

\author{Yan Zhuang\inst{1}$^{\star}$ \and Minheng Chen\inst{1}$^{\star}$ \and
Chao Cao\inst{1} \and Tong Chen\inst{1} \and Jing Zhang\inst{1} \and Xiaowei Yu\inst{1} \and Yanjun Lyu\inst{1} \and 
Lu Zhang\inst{2} \and Tianming Liu\inst{3} \and Dajiang Zhu\inst{1}}
\authorrunning{Y. Zhuang et al.}
\institute{Department of Computer Science and Engineering, University of Texas at Arlington, USA \and
    Department of Computer Science, Indiana University Indianapolis, USA \and
    School of Computing, University of Georgia, USA
    }
    
\maketitle              
\begin{abstract}
Understanding the structural and functional organization of the human brain requires a detailed examination of cortical folding patterns, among which the three-hinge gyrus (3HG) has been identified as a key structural landmark. GyralNet, a network representation of cortical folding, models 3HGs as nodes and gyral crests as edges, highlighting their role as critical hubs in cortico-cortical connectivity. However, existing methods for analyzing 3HGs face significant challenges, including the sub-voxel scale of 3HGs at typical neuroimaging resolutions, the computational complexity of establishing cross-subject correspondences, and the oversimplification of treating 3HGs as independent nodes without considering their community-level relationships. To address these limitations, we propose a fully differentiable subnetwork partitioning framework that employs a spectral modularity maximization optimization strategy to modularize the organization of 3HGs within GyralNet. By incorporating topological structural similarity and DTI-derived connectivity patterns as attribute features, our approach provides a biologically meaningful representation of cortical organization. Extensive experiments on the Human Connectome Project (HCP) dataset demonstrate that our method effectively partitions GyralNet at the individual level while preserving the community-level consistency of 3HGs across subjects, offering a robust foundation for understanding brain connectivity.

\end{abstract}
\renewcommand{\thefootnote}{\fnsymbol{footnote}}
\footnotetext[1]{The two authors contributed equally to this work.}
\renewcommand{\thefootnote}{\arabic{footnote}}
\section{Introduction}
GyralNet is a network-like cortical folding organization that represents the overall gyral morphology of the whole brain, in which three-hinge gyrus (3HG) act as nodes and gyral crests act as edges~\cite{chen2017gyral,li2010gyral}.
Recent studies have characterized 3HG as a finer-scale cortical folding pattern that remains evolutionarily conserved across multiple primate species and exhibits remarkable stability regardless of individual variability, population differences, or pathological states~\cite{li2017commonly,zhang2018exploring}.
This cortical folding architecture demonstrates high intra-species consistency while exhibiting inter-individual variability, characterized by distinct features such as the greatest cortical thickness~\cite{li2010gyral}, increased DTI-derived fiber density~\cite{ge2018denser}, and enhanced connectivity heterogeneity across both structural and functional domains compared to other gyral regions~\cite{zhang2020cortical}.
These properties indicate that 3HGs serve as critical hubs within the cortico-cortical connectivity framework and play a pivotal role in the global structural and functional architecture of the human brain~\cite{liu2022optimized}.

Cortical arealization has become a fundamental concept in systems neuroscience, where a cortical area is defined as a fixed location with its own unique inputs, outputs, and internal organization~\cite{petersen2024principles}. 
While voxel-level image analysis is possible and common, targeting areas composed of collections of voxels is often more useful for scientific inquiry, as these represent primary underlying neurobiological objects of interest~\cite{wig2011concepts}.
The significant role of 3HGs in brain functional connectivity~\cite{zhang2023joint}, coupled with their anatomical distinctiveness that remains unaffected by uncertainty in brain region boundary definitions~\cite{he2024brain}, makes them ideal candidates for constructing connectomes based on cortical analysis. A recent study revealed that a finer-scale brain connectome based on 3HGs can better capture the intricate patterns of Alzheimer's disease~\cite{lyu20204mild}. 
However, due to the limitations of current neuroimaging data resolution, this folding pattern is nearly at a sub-voxel scale at typical effective resolutions (slightly greater than 1mm), introducing considerable noise and interference in studies based on individual 3HGs.
This greatly complicates the frameworks employed in previous studies that attempt to establish consistent correspondences across subjects based on a single 3HG~\cite{cao2024enhance,zhang2024learning,zhang2023cortex2vector}, making the process both cumbersome and computationally demanding.
In addition, prior studies treat all 3HGs as “ordinary” nodes before establishing cross-subject correspondences, without accounting for their potential community-level relationships, as observed in~\cite{zhang2023joint}.
This may result in a failure to capture essential functional interactions and could potentially lead to oversimplified or inaccurate representations of brain networks.

In this paper, to overcome the limitations associated with analyzing individual 3HGs, we propose a fully differentiable subnetwork partitioning framework that employs a spectral modularity maximization optimization strategy to further partition 3HGs within GyralNet. This framework leverages topological structural similarity and DTI-derived structural connectivity patterns as attribute features for each 3HG, enabling a more comprehensive and biologically meaningful representation of cortical organization.
Extensive experiments conducted on the Human Connectome Project (HCP) dataset demonstrate that our method not only effectively partitions GyralNet at the individual level but also preserves the community-level consistency of 3HGs across subjects, as evidenced by both quantitative and qualitative results.
\section{Methodology}
\subsection{Preliminaries}
\noindent\textbf{Problem definition.}
Since GyralNet is a network-like cortical folding organization, and each 3HG serves as a node in the network. So it can be represented as an undirected and unweighted graph: $G = (V, E)$, where $|V|=n$ and $|E|=m$. 
 We denote the \( n \times n \) adjacency matrix of \( G \) by \( \mathbf{A} \), where \( \textbf{A}_{ij} = 1 \) iff \( \{v_i, v_j\} \in E \). The degree of \( v_i \) is its number of connections \( d_i := \sum_{j=1}^{n} \textbf{A}_{ij} \). 
Our goal is to evaluate the quality of the graph partitioning function \( \mathcal{F} : V \mapsto \{1, \dots, k\} \), which partitions the node set \( V \) into 
$k$ subsets $V_i=\{v_j:\mathcal{F}(v_j)=i\}$, taking into account both the structural topology of the graph and the intrinsic properties of the nodes $ \mathbf{X} \in \mathbb{R}^{n \times s}$.

\noindent\textbf{Graph modularity.}
Graph modularity~\cite{newman2006modularity} is one of the most widely used and highly effective metrics for graph clustering in scientific research. It quantifies the strength of a network's division into distinct communities by measuring the deviation of intra-community edge density from what would be expected in a randomly connected network. Mathematically, modularity is defined as:
\begin{equation}
\label{eq::modularity}
    Q = \frac{1}{2m} \sum_{i,j} \left[ \textbf{A}_{ij} - \frac{d_i d_j}{2m} \right] \delta(c_i, c_j)
\end{equation}
where  $\delta(c_i, c_j) = 1$ if i and j are in the same cluster and 0 otherwise.
A higher modularity score indicates a stronger community structure, where connections within clusters are denser than would be expected by chance.
\begin{figure}[htb]

  \centering
  \centerline{\includegraphics[width=\linewidth]{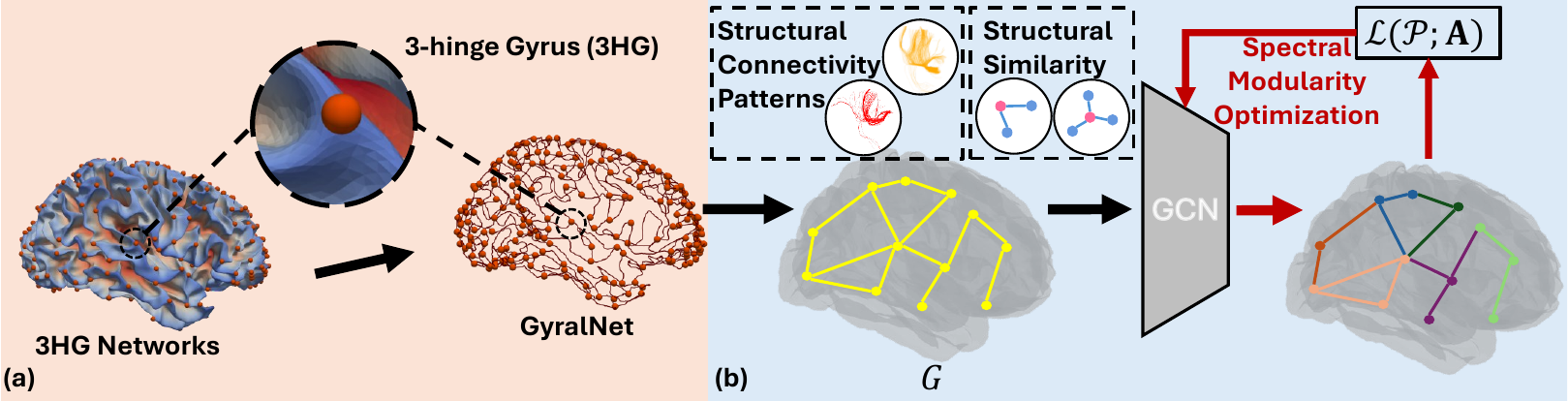}}
%
\caption{Overall architecture of the proposed method. 
(a) The folding pattern of the cerebral cortex organized by 3HGs. (b) An illustration of the differentiable spectral modularity optimization, the structural representation of each 3HG integrates both structural similarity and structural connectivity patterns.}
\label{fig:overall}
\end{figure}
\subsection{Structural Representation of 3-Hinge Gyrus}
The structural representation of 3HG can be divided into two components: the feature vector $X_s$, which captures the topological similarity among 3HG nodes, and the feature vector $X_c$, which characterizes the structural connection patterns within 3HGs.
Thus, the node attributes of the i-th 3HG in GyralNet can be expressed as: $X^i=\{X^i_s,X^i_c\}\in \mathbb{R}^{1 \times s}$.
The structural similarity between two nodes is determined by comparing the topological structure of the nodes and their neighbors~\cite{ribeiro2017struc2vec}.
Following the approach in~\cite{chen2024using,zhang2023cortex2vector}, we represent the structural information of a 3HG using the degrees of its neighboring nodes. 
Specifically, we define the feature vector as:
\begin{equation}
X^i_s=\langle S\odot I F\rangle _{i}\in \mathbb{R}^{1\times r}
\end{equation}
where $S \in \mathbb{R}^{n\times n}$ represents the structural similarity of the entire graph, $F \in \mathbb{R}^{n\times r}$ the one-hot ROI label of each 3HG obtained from a specific atlas ( $r$ being the number of ROIs in the atlas), $\langle \cdot \rangle _{i}$ denotes the i-row of the matrix and $\odot$ indicates the Hadamard product. And the structural similarity between two 3HGs $u$ and $v$ can be expressed as follows:
\begin{equation}\label{eq1}
S (u,v)=e^{-w(u,v)}
\end{equation}
$w(u,v)$ is a similarity measure between two sequences obtained by calculating the neighborhood of two 3HGs in order of degree based on the dynamic time warping algorithm~\cite{muller2007dynamic}.
As for $X_c$, we utilize Trace-map~\cite{zhu2013dicccol,zhu2011discovering} to characterize the structural connection pattern of 3HG. 
Trace-map is a morphological feature of fiber bundles designed to describe cortical landmarks. It captures fiber morphology by projecting each fiber segment onto a sphere and then applying histogram-based statistics to quantify the distribution of fiber segments across different regions of the sphere.
\subsection{Network Architecture}
We employ a two-layer Graph Convolutional Network (GCN) to construct the graph partitioning function $\mathcal{F}(\cdot)$.
Similar to the design in~\cite{tsitsulin2023graph}, we added learnable skip connections $ \mathbf{W}_{\text{skip}}\in \mathbb{R}^{s\times s}$ to each layer of GCN and replaced the activation function from ReLU to SeLU for better convergence.
The input to the $t+1$-th layer of the GCN consists of the feature matrix $\mathbf{X}_t \in \mathbb{R}^{n\times s}$ and the normalized adjacency matrix $\Tilde{\textbf{A}}=\textbf{D}^{-1/2}\textbf{A}\textbf{D}^{-1/2}$, producing the following output:
\begin{equation}
    \mathbf{X}_{t+1} = \text{SeLU} (\tilde{\mathbf{A}} \mathbf{X}_{t} \mathbf{W} + \mathbf{X}_{t} \mathbf{W}_{\text{skip}})
\end{equation}
The output of the network passes through a softmax function to obtain a soft assignment probability $\mathcal{P}\in \mathbb{R}^{n\times k}$ (shown in Eq.~\ref{eq::network}), which makes the assignment process differentiable .
\begin{equation}
\label{eq::network}
  \mathcal{P}= \text{softmax}(\text{GCN}(\tilde{\mathbf{A}}, \mathbf{X}))
\end{equation}
\subsection{Spectral Modularity Optimization}
Maximizing modularity has been proven to be NP-hard. While certain relaxation-based algorithms can be used to approximate a solution, these methods operate solely on graph structures, making their adaptation to property graphs non-trivial.
The proposed differentiable spectral modularity optimization seeks to maximize an effective spectral relaxation of modularity~\cite{newman2006finding}. 
Let $\mathbf{d}\in \mathbb{R}^{n\times 1}$ denote the degree vector. Under the given conditions, the modularity can be reformulated as:
\begin{equation}
    Q = \frac{1}{2m} \operatorname{Tr}(\mathcal{P}^{\top} \mathbf{B} \mathcal{P}), \text{where}\: \mathbf{B} = \mathbf{A} - \frac{\mathbf{d} \mathbf{d}^{\top}}{2m}
\end{equation}
Thus, our optimization objective function is to find an optimal $\hat{\mathcal{P}}$ that satisfies:
\begin{equation}
    \hat{\mathcal{P}}=\arg\min_{\mathcal{P}}-\frac{1}{2m} \operatorname{Tr}(\mathcal{P}^{\top} \mathbf{B} \mathcal{P})
\end{equation}
To prevent our gradient-based optimization from becoming trapped in local minima, we incorporate collapse regularization~\cite{tsitsulin2023graph}, which serves as a constraint for spectral clustering in both the min-cut objective and the modularity objective. This regularization prevents trivial splits while ensuring that the optimization of the main objective remains unaffected.
\begin{equation}
\label{eq::loss}
    \mathcal{L}(\mathcal{P}; \mathbf{A}) = 
- \frac{1}{2m} \operatorname{Tr}(\mathcal{P}^{\top} \mathbf{B} \mathcal{P})
+ \frac{\sqrt{k}}{n} \left\| \sum_{i} \mathcal{P}_i^{\top} \right\|_F - 1
\end{equation}
The loss function of our framework is shown in Eq~\ref{eq::loss}, where $\left\|\cdot\right\|_F$ is the Frobenius norm.
The proposed framework iteratively partitions subnetworks within the GyralNet, which is extracted from a human hemisphere, until the loss function converges.
\section{Experiment}
\subsection{Experimental Settings}
\subsubsection{Dataset.}
In this study, we utilize T1-weighted structural MRI (sMRI) and diffusion tensor imaging (DTI) data from 1,056 subjects in the Human Connectome Project (HCP) S1200 release~\cite{van2013wu}. 
The sMRI data undergoes the same pre-processing procedures as described in **.
Specifically, the pre-processing pipeline includes brain skull removal, tissue segmentation, and cortical surface reconstruction using the FreeSurfer package. For DTI data, eddy current correction is performed using FSL, followed by fiber tracking with DSI Studio~\cite{yeh2013deterministic}.
After registering the sMRI data to DTI space, the Destrieux Atlas is applied to label regions of interest (ROIs) on the reconstructed white matter surface. 
The GyralNet is then extracted from the cortical surface following the pipeline described in~\cite{chen2017gyral}. 
Additionally, Trace-map is computed for each 3HG using the approach outlined in~\cite{zhu2013dicccol}.

\subsubsection{Implementation details.}
The differentiable optimization process is performed using the Adam optimizer with a learning rate of 1e-3. 
The network consists of two GCN layers with dimensions (128, 64). 
The maximum number of iterations is set to 1500, and when the variance of the loss function over the last 10 iterations falls below 1e-3, it is considered to have early converged.
The entire framework is implemented in PyTorch. All experiments are conducted on a desktop equipped with an NVIDIA RTX 6000 Ada GPU and a 3.6 GHz Intel CPU.
\begin{table}[h!]
    \centering
    \caption{Quantitative results of the proposed method. Comparison with baselines that rely solely on structural similarity or structural connectivity patterns.}
    \label{tab:results}
    \renewcommand{\arraystretch}{1.2}
    \setlength{\tabcolsep}{6pt}
    \begin{tabular}{c c cc c cccc c}
        \toprule
        & \multicolumn{3}{c}{\textbf{Ours}} & \multicolumn{3}{c}{\textbf{Similarity Only}} & \multicolumn{3}{c}{\textbf{Connectivity Only}} \\
        \cmidrule(lr){2-4} \cmidrule(lr){5-7} \cmidrule(lr){8-10}
         & $\mathcal{C} \downarrow$ & $Q \uparrow$ & CS$\uparrow$& $\mathcal{C} \downarrow$ & $Q \uparrow$  & CS$\uparrow$& $\mathcal{C} \downarrow$ & $Q \uparrow$ & CS$\uparrow$ \\
        \Xhline{1pt}
        \multicolumn{1}{l|}{$k=4$ } & \textbf{11.6}  & 59.5& \multicolumn{1}{c|}{\textbf{71.9}} & \textbf{14.7}& 57.7  &  \multicolumn{1}{c|}{\textbf{48.6}} & \textbf{13.9}& 53.5  &\textbf{69.9} \\
        \multicolumn{1}{l|}{$k=8$} & 14.4 & 62.9 &\multicolumn{1}{c|}{57.0}  & 18.3 & 60.1 &\multicolumn{1}{c|}{32.8}  & 17.7 & 56.7 &57.5  \\
        \multicolumn{1}{l|}{$k=12$} & 16.2& 64.8&\multicolumn{1}{c|}{54.6}   &20.6 &61.5  & \multicolumn{1}{c|}{27.4}   & 20.2& 58.4&50.7 \\
        \multicolumn{1}{l|}{$k=16$} & 17.6 &65.9 & \multicolumn{1}{c|}{48.1}  &22.2 &62.2 &\multicolumn{1}{c|}{24.2}    & 22.2& \textbf{59.1} &48.9 \\
        \multicolumn{1}{l|}{$k=20$} & 18.7 & 66.6&\multicolumn{1}{c|}{45.9}  &23.5& 62.6&  \multicolumn{1}{c|}{21.7}  & 24.0 & \textbf{59.1}&44.3  \\
        \multicolumn{1}{l|}{$k=24$} & 19.7 &67.0 & \multicolumn{1}{c|}{42.6}  &24.6 &\textbf{62.9}&\multicolumn{1}{c|}{20.6}    & 25.6 & 58.9 &42.5   \\
        \multicolumn{1}{l|}{$k=32$} & 20.8&\textbf{67.1} & \multicolumn{1}{c|}{40.1}  &25.7& \textbf{62.9} & \multicolumn{1}{c|}{18.6}  & 27.1 & 58.1&41.7  \\
        \multicolumn{1}{l|}{$k=40$} & 21.8 &67.0 &\multicolumn{1}{c|}{37.0}   & 26.8&  62.8 &\multicolumn{1}{c|}{17.0}   & 28.4& 57.3&40.2   \\
        \multicolumn{1}{l|}{$k=50$} & 22.9 &66.7 & \multicolumn{1}{c|}{35.0}  & 27.8& 62.5& \multicolumn{1}{c|}{15.4}   & 29.5& 56.6 &39.4 
        \\
        \bottomrule
    \end{tabular}

\end{table}
\subsubsection{Evaluation metrics.}
We employed graph-based clustering metrics, including cluster conductance and graph modularity, to evaluate the community structure of the proposed subnetwork partitioning method within the subject. Cluster conductance~\cite{yang2012defining} measures the extent to which a community is connected to the rest of the graph, and it is computed as follows:
\begin{equation}
\label{eq::conductance}
\mathcal{C} = \frac{1}{k} \sum_{i=1}^{k} 
\frac{|\{(u,v) \mid u \in V_i, v \notin V_i\}|}
{\min\left(\sum_{u \in V_i} d_u, \sum_{v \in V \setminus V_i} d_v \right)}
\end{equation}
Graph modularity, on the other hand, assesses the effectiveness of the overall partitioning in capturing the community structure of the graph, with its formula presented in Eq.~\ref{eq::modularity}. 
In addition, we assess the cosine similarity(CS) of the structural representation features of the corresponding subnetworks across subjects to evaluate the consistency of subnetwork partitioning. The correspondence of subnetworks across subjects is established by first computing the Pearson correlation coefficient between their localization descriptors. The \textbf{localization descriptor} of a 3HG subnetwork is defined as the accumulated one-hot vector representation of ROIs corresponding to each 3HG it contains. The subnetworks are then matched based on these localization descriptors using the classic Kuhn-Munkres algorithm.
To enhance readability and facilitate comparison, all metrics were multiplied by 100 where applicable.

\subsection{Results}
\begin{figure}[h!]

  \centering
  \centerline{\includegraphics[width=\linewidth]{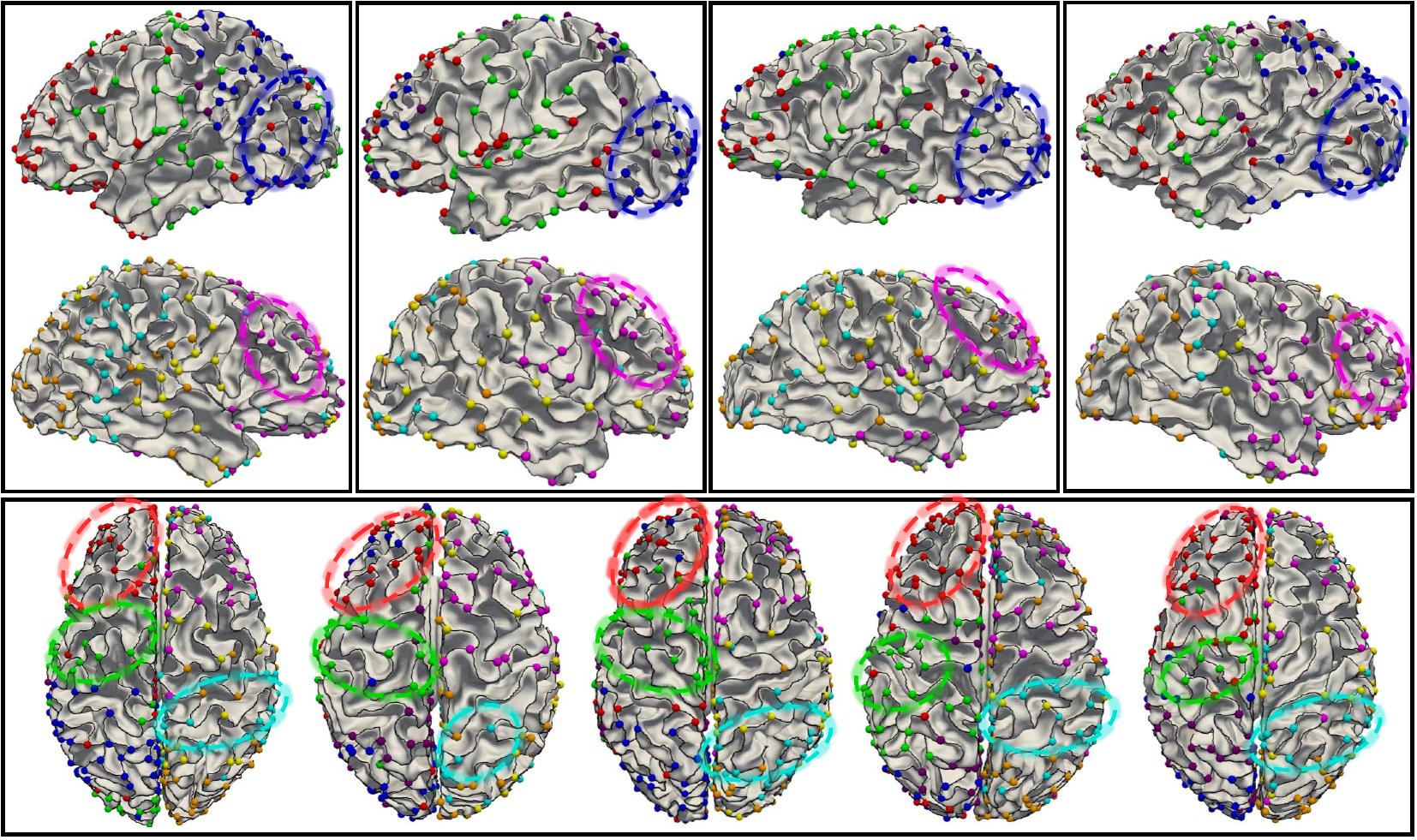}}
%
\caption{Partitioned GyralNets ($k=4$) from multiple subjects are shown in lateral (top: left hemisphere, middle: right hemisphere) and superior (bottom) views. Each block represents a subject, with distinct colors indicating subnetworks. Elliptical outlines highlight consistently observed subnetworks across subjects.}
\label{fig:Visualization}
\end{figure}
We report the conductance, modularity, and cosine similarity of the corresponding subnetwork features across subjects for different values of $k$, as shown in Table~\ref{tab:results}.
It is important to note that since subnetwork partitioning is performed separately on the left and right hemispheres of each subject, so the final result consists of $2k$ 3HGs communities per individual.
In addition, we also compared the proposed framework using only structural similarity and only structural connectivity patterns as node attributes for subnetwork partitioning.
Our framework processes each individual in 5 to 10 seconds, depending on the number of 3HGs and the chosen $k$
We observed that as $k$ increases, the conductance of the graph gradually rises, while modularity initially increases before eventually declining. This indicates that finer subnetwork partitioning of 3HGs does not necessarily lead to better results, as excessive fragmentation can disrupt the overall network structure, whereas overly coarse partitioning may fail to capture meaningful community organization.
Therefore, selecting an appropriate $k$ value is crucial. Based on the experimental results, the optimal range for $k$ appears to be approximately between 16 and 32, ensuring a balance between modularity and network structure preservation.
The proposed method consistently achieves lower conductance and higher modularity across all values of $k$, indicating that the identified subnetworks are more well-separated and the resulting communities are better structured compared to those obtained using partial node attributes.
This further demonstrates that the joint representation of structural similarity and structural connectivity patterns of 3HGs is effective in accurately identifying 3HG subnetworks.
The CS of subnetworks across subjects, obtained using the full 3HG structure representation for subnetwork partitioning, is the highest, confirming its effectiveness. Additionally, as the number of communities increases, CS gradually decrease with increasing $k$, suggesting that finer subnetwork partitions lead to greater inter-individual variability. This implies that while more granular partitioning may capture localized structural differences, it also reduces the stability of subnetwork alignment across subjects.
\begin{figure}[htb]
  \centering
  \centerline{\includegraphics[width=\linewidth]{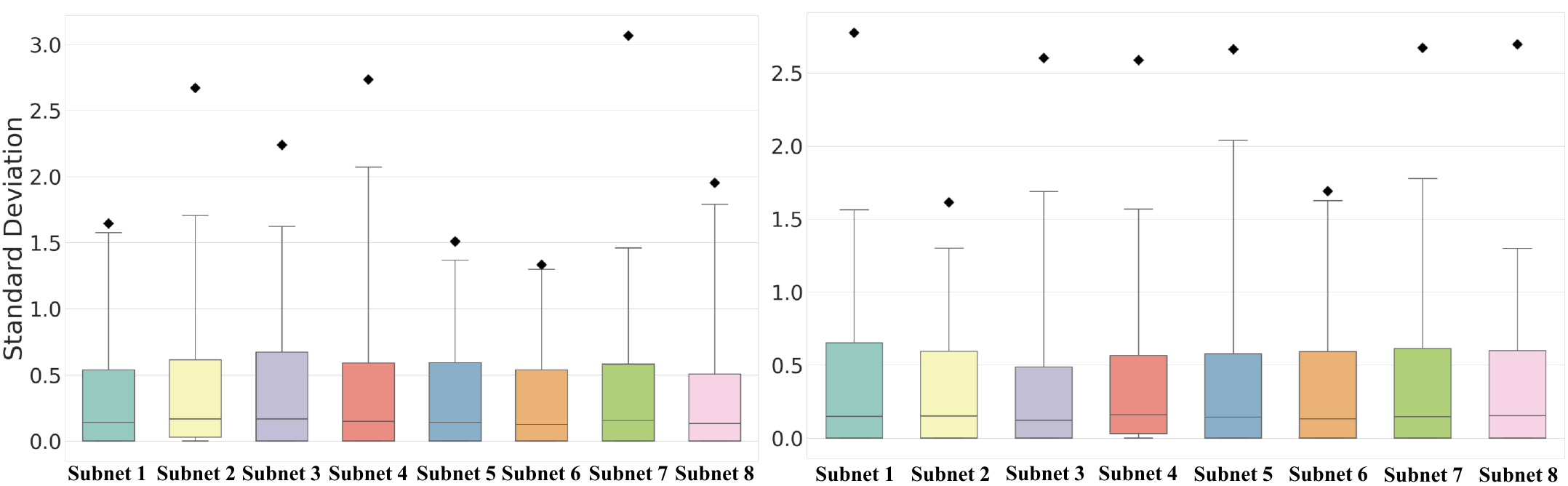}}
%
\caption{The figures on the left and right show the mean of the standard deviation of the localization descriptor ($k=8$) in the left and right hemispheres across all subjects from HCP dataset, respectively.}
\label{fig:quantitative}
\end{figure}
In Figure~\ref{fig:quantitative}, we present the mean standard deviation of the localization descriptor across all individuals for $k=8$. The results show that the average standard deviation of each subnetwork across the dataset is approximately 0.2, indicating a high level of consistency in subnetwork distribution across subjects.
Figure~\ref{fig:Visualization} visualizes the subnetwork partitioning, where different subnetworks are highlighted using ellipses of distinct colors in each brain. Since brain connectivity is not strictly constrained by cortical folding patterns, the spatial locations of the same subnetwork can vary across individuals. Moreover, adjacent 3HGs within an individual do not necessarily belong to the same subnetwork, as structural connectivity is primarily mediated by white matter pathways rather than direct cortical proximity. As a result, distant 3HGs may be part of the same community, reflecting the complex and non-local nature of brain network.
\section{Conclusion}
This paper proposes a fully differentiable subnetwork partitioning framework that leverages spectral modularity maximization to modularize the organization of 3HGs within GyralNet. 
By integrating topological structural similarity and DTI-derived connectivity patterns as attribute features, our method provides a more biologically meaningful representation of cortical organization.
Experimental results on the HCP dataset demonstrate that our approach effectively partitions GyralNet at the individual level while maintaining community-level consistency across subjects.
Future work will focus on developing a more comprehensive cross-individual subnetwork correspondence strategy and utilizing community-level 3HG subnetworks to investigate brain function and their potential role in identifying neurological disorders.
\bibliographystyle{splncs04}
\bibliography{refs}

\begin{thebibliography}{10}
\providecommand{\url}[1]{\texttt{#1}}
\providecommand{\urlprefix}{URL }
\providecommand{\doi}[1]{https://doi.org/#1}

\bibitem{cao2024enhance}
Cao, C., Yu, X., Zhang, L., Chen, T., Lyu, Y., Liu, T., Zhu, D.: Enhancing group-wise consistency in 3-hinge gyrus matching via anatomical embedding and structural connectivity optimization. In: 2024 IEEE International Symposium on Biomedical Imaging (ISBI). pp.~1--5 (2024)

\bibitem{chen2017gyral}
Chen, H., Li, Y., Ge, F., Li, G., Shen, D., Liu, T.: Gyral net: a new representation of cortical folding organization. Medical image analysis  \textbf{42},  14--25 (2017)

\bibitem{chen2024using}
Chen, M., Cao, C., Chen, T., Zhuang, Y., Zhang, J., Lyu, Y., Yu, X., Zhang, L., Liu, T., Zhu, D.: Using structural similarity and kolmogorov-arnold networks for anatomical embedding of 3-hinge gyrus. arXiv preprint arXiv:2410.23598  (2024)

\bibitem{ge2018denser}
Ge, F., Li, X., Razavi, M.J., Chen, H., Zhang, T., Zhang, S., Guo, L., Hu, X., Wang, X., Liu, T.: Denser growing fiber connections induce 3-hinge gyral folding. Cerebral Cortex  \textbf{28}(3),  1064--1075 (2018)

\bibitem{he2024brain}
He, Z., Zhang, T., Wang, Q., Zhang, S., Cao, G., Liu, T., Zhao, S., Jiang, X., Guo, L., Yuan, Y., et~al.: Brain functional gradients are related to cortical folding gradient. Cerebral Cortex  \textbf{34}(11),  bhae453 (2024)

\bibitem{li2010gyral}
Li, K., Guo, L., Li, G., Nie, J., Faraco, C., Cui, G., Zhao, Q., Miller, L.S., Liu, T.: Gyral folding pattern analysis via surface profiling. NeuroImage  \textbf{52}(4),  1202--1214 (2010)

\bibitem{li2017commonly}
Li, X., Chen, H., Zhang, T., Yu, X., Jiang, X., Li, K., Li, L., Razavi, M.J., Wang, X., Hu, X., et~al.: Commonly preserved and species-specific gyral folding patterns across primate brains. Brain structure and function  \textbf{222},  2127--2141 (2017)

\bibitem{liu2022optimized}
Liu, S., Ge, F., Zhao, L., Wang, T., Ni, D., Liu, T.: Nas-optimized topology-preserving transfer learning for differentiating cortical folding patterns. Medical image analysis  \textbf{77},  102316 (2022)

\bibitem{lyu20204mild}
Lyu, Y., Zhang, L., Yu, X., Cao, C., Liu, T., Zhu, D.: Mild cognitive impairment classification using a novel finer-scale brain connectome. In: 2024 IEEE International Symposium on Biomedical Imaging (ISBI). pp.~1--5 (2024)

\bibitem{muller2007dynamic}
M{\"u}ller, M.: Dynamic time warping. Information retrieval for music and motion pp. 69--84 (2007)

\bibitem{newman2006finding}
Newman, M.E.: Finding community structure in networks using the eigenvectors of matrices. Physical Review E—Statistical, Nonlinear, and Soft Matter Physics  \textbf{74}(3),  036104 (2006)

\bibitem{newman2006modularity}
Newman, M.E.: Modularity and community structure in networks. Proceedings of the national academy of sciences  \textbf{103}(23),  8577--8582 (2006)

\bibitem{petersen2024principles}
Petersen, S.E., Seitzman, B.A., Nelson, S.M., Wig, G.S., Gordon, E.M.: Principles of cortical areas and their implications for neuroimaging. Neuron  (2024)

\bibitem{ribeiro2017struc2vec}
Ribeiro, L.F., Saverese, P.H., Figueiredo, D.R.: struc2vec: Learning node representations from structural identity. In: Proceedings of the 23rd ACM SIGKDD international conference on knowledge discovery and data mining. pp. 385--394 (2017)

\bibitem{tsitsulin2023graph}
Tsitsulin, A., Palowitch, J., Perozzi, B., M{\"u}ller, E.: Graph clustering with graph neural networks. Journal of Machine Learning Research  \textbf{24}(127),  1--21 (2023)

\bibitem{van2013wu}
Van~Essen, D.C., Smith, S.M., Barch, D.M., Behrens, T.E., Yacoub, E., Ugurbil, K., Consortium, W.M.H., et~al.: The wu-minn human connectome project: an overview. Neuroimage  \textbf{80},  62--79 (2013)

\bibitem{wig2011concepts}
Wig, G.S., Schlaggar, B.L., Petersen, S.E.: Concepts and principles in the analysis of brain networks. Annals of the New York Academy of Sciences  \textbf{1224}(1),  126--146 (2011)

\bibitem{yang2012defining}
Yang, J., Leskovec, J.: Defining and evaluating network communities based on ground-truth. In: 2012 IEEE 12th International Conference on Data Mining. pp. 745--754 (2012)

\bibitem{yeh2013deterministic}
Yeh, F.C., Verstynen, T.D., Wang, Y., Fern{\'a}ndez-Miranda, J.C., Tseng, W.Y.I.: Deterministic diffusion fiber tracking improved by quantitative anisotropy. PloS one  \textbf{8}(11),  e80713 (2013)

\bibitem{zhang2024learning}
Zhang, L., Wu, Z., Yu, X., Lyu, Y., Wu, Z., Dai, H., Zhao, L., Wang, L., Li, G., Wang, X., et~al.: Learning lifespan brain anatomical correspondence via cortical developmental continuity transfer. Medical Image Analysis p. 103328 (2024)

\bibitem{zhang2023cortex2vector}
Zhang, L., Zhao, L., Liu, D., Wu, Z., Wang, X., Liu, T., Zhu, D.: Cortex2vector: anatomical embedding of cortical folding patterns. Cerebral Cortex  \textbf{33}(10),  5851--5862 (2023)

\bibitem{zhang2023joint}
Zhang, S., Wang, R., Kang, Y., Yu, S., Hu, H., Zhang, H.: Joint representation of functional and structural profiles for identifying common and consistent 3-hinge gyral folding landmark. In: International Conference on Medical Image Computing and Computer-Assisted Intervention. pp. 163--172. Springer (2023)

\bibitem{zhang2018exploring}
Zhang, T., Chen, H., Razavi, M.J., Li, Y., Ge, F., Guo, L., Wang, X., Liu, T.: Exploring 3-hinge gyral folding patterns among hcp q3 868 human subjects. Human brain mapping  \textbf{39}(10),  4134--4149 (2018)

\bibitem{zhang2020cortical}
Zhang, T., Li, X., Jiang, X., Ge, F., Zhang, S., Zhao, L., Liu, H., Huang, Y., Wang, X., Yang, J., et~al.: Cortical 3-hinges could serve as hubs in cortico-cortical connective network. Brain imaging and behavior  \textbf{14},  2512--2529 (2020)

\bibitem{zhu2013dicccol}
Zhu, D., Li, K., Guo, L., Jiang, X., Zhang, T., Zhang, D., Chen, H., Deng, F., Faraco, C., Jin, C., et~al.: Dicccol: dense individualized and common connectivity-based cortical landmarks. Cerebral cortex  \textbf{23}(4),  786--800 (2013)

\bibitem{zhu2011discovering}
Zhu, D., Zhang, D., Faraco, C., Li, K., Deng, F., Chen, H., Jiang, X., Guo, L., Miller, L.S., Liu, T.: Discovering dense and consistent landmarks in the brain. In: Information Processing in Medical Imaging: 22nd International Conference, IPMI 2011, Kloster Irsee, Germany, July 3-8, 2011. Proceedings 22. pp. 97--110. Springer (2011)

\end{thebibliography}
%




\end{document}